\documentclass[]{llncs}

\usepackage{float}
\floatstyle{boxed}
\newfloat{figure}{h}{ext}
\floatname{figure}{Figure}
\usepackage{graphicx} 
\usepackage{amssymb}

\newcommand{\Prop}{{\it Prop}}

\newcommand{\nat}{{\bf N}}

\newcommand{\rimp}{\Rightarrow}
\newcommand{\dimp}{\Leftrightarrow}

\newcommand{\I}{{\cal I}}
\newcommand{\R}{{\cal R}}

\newcommand{\Next}{X}

\newcommand{\powerset}[1]{{\cal P}(#1)}

\newcommand{\be}{\begin{enumerate}}
\newcommand{\ee}{\end{enumerate}}

\newcommand{\dlvrd}{{\tt dlvrd}}
\newcommand{\rr}{{\tt rr}}
\newcommand{\kc}{{\tt kc}}
\newcommand{\said}{{\tt said}}
\newcommand{\keyleft}{{\tt keyleft}}
\newcommand{\keyright}{{\tt keyright}}
\newcommand{\sender}{{\tt sender}}
\newcommand{\msg}{{\tt message}} 
\newcommand{\rcvd}{{\tt rcvd}}
\newcommand{\conflict}{{\tt conflict}}
\newcommand{\conflictfree}{{\tt conflict\mbox{-}free}}
\newcommand{\slotrequest}{{\tt slot\mbox{-}request}}

\newcommand{\false}{0} 
\newcommand{\true}{1} 

\newcommand{\commentout}[1]{}

\newcommand{\versiondate}{April 20, 2010}

\title{Epistemic Model Checking for Knowledge-Based Program Implementation:
an Application to Anonymous Broadcast\thanks{Version of \versiondate. This material is based on research sponsored by the Air Force Research Laboratory, under agreement number FA2386-09-1-4156. The U.S. Government is authorized to reproduce and distribute reprints for Governmental purposes notwithstanding any copyright notation thereon. The views and conclusions contained herein are those of the authors and should not be interpreted as necessarily representing the official policies or endorsements, either expressed or implied, of the Air Force Research Laboratory or the U.S. Government.}}  

\author{Omar I. Al-Bataineh \and Ron van der Meyden\\
\{omara,meyden\}@cse.unsw.edu.au
}
\institute{School of Computer Science and Engineering, \\ 
University of New South Wales}

\begin{document}

\maketitle

\begin{abstract}
Knowledge-based programs provide an abstract level of description of
protocols in which agent actions are related to their states of
knowledge.  The paper describes how epistemic model checking
technology may be applied to discover and verify concrete
implementations based on this abstract level of description. The
details of the implementations depend on the specific context of use
of the protocol. The knowledge-based approach enables the
implementations to be optimized relative to these conditions of use.
The approach is illustrated using extensions of the Dining
Cryptographers protocol, a security protocol for anonymous broadcast.
\end{abstract}

\section{Introduction}

In distributed systems, we generally would like agent's actions to depend upon the information that they have. However, the way that information flows in such systems can be quite complex. It has been proposed to address this complexity by the use of formal logics of knowledge \cite{FHMVbook}.

 In particular, {\em knowledge based programs} have been proposed as a level of abstraction that directly captures the relationship between an agent's knowledge and its actions, by allowing branching statements to contain formulas of the modal logic of knowledge, expressing what the agent knows about the global state of the system. This has several advantages. By focusing on what information is required, rather than how it is encoded, knowledge-based programs can be more intuitive and more easily verified to be correct. They can also provide a common description that is independent of assumptions such as the failure modes of communication channels in the system. Finally, knowledge-based programs lead us to implementations that are {\em optimal} in their use of information, in the sense
that agents do not overlook opportunities to use relevant information that is available in their local states. 
 
A cost of the abstraction that knowledge-based programs provide, is that they 
are more like specifications than concrete programs, so cannot be directly executed. To obtain an executable program, is necessary to replace the tests for knowledge in the knowledge based program by equivalent concrete predicates of the agent's local state. Because of the complexity of information flow in distributed systems, such concrete predicates can be difficult to find. To date, this task has generally been carried out by pencil and paper reasoning. Perhaps for this reason, there remain only a handful of worked out examples of the development of concrete implementations of knowledge-based programs (e.g., \cite{BaukusM04,DM86,DM90,Had87,HZ92}). 
 
The difficulty can be addressed through the use of model checking technology for the logic of knowledge. Model checkers are systems that take as input a formal model of a system, together with a specification, and determine whether that specification is satisfied by the model 
 \cite{CGP99}. 
 The specification language used in model checkers is generally a form of temporal logic, but in recent years work has begun on the development of model checkers based on logics of knowledge \cite{mck,mcmas,demo}. We describe a methodology for the use of this latter class of model checkers to the development of implementations of knowledge based programs. The methodology is partially automated. It assists users in finding a concrete predicates that are equivalent to the knowledge conditions in a knowledge-based program by means of an iterative process, in which automatically computed counterexamples to a user's guess for the concrete predicate are used by the user to construct an improved concrete predicate, until one is found that is equivalent to the knowledge condition. 
 
We illustrate the methodology by means of an example in which we use the epistemic model checker MCK \cite{mck}, to develop concrete implementations of a knowledge-based program for anonymous broadcast, based on multiple rounds of Chaum's Dining Cryptographers Protocol \cite{chaum}. 
 
The Dining Cryptographers Protocol enables a message to be broadcast anonymously, under the assumption that only one agent is attempting to send broadcast a message. The objective of the extension that we consider is to remove this assumption, so that any number of agents may broadcast their messages anonymously. One of the main difficulties in this is that, since agents operate independently, it is possible for simultaneous broadcasts to interfere with each other, causing a failure in the transmission. Thus, a key issue is to enable the agents to detect conflicts in the transmission, and to respond appropriately when a conflict is detected.
 
In our analysis, we express the expected behaviour using a knowledge based program that conditions the agent's actions on whether it knows that there is a conflict. We then use our model checking supported methodology to identify exactly the concrete conditions under which an agent knows whether there is a conflict. These conditions turn out to have a surprising level of complexity. In particular, we find that these conditions can differ, depending on the assumptions that we make about the number of agents wishing to broadcast. 
  
 Our approach leads to the discovery (assisted by automation) of a number of subtleties concerning the protocol that, to our knowledge, have not been previously noticed. In particular, we find that it is possible for agents to detect conflicts (or lack of conflict) in some quite unexpected situations. Moreover, we discover situations where, even though the protocol terminates, an agent cannot be sure that its message has been successfully transmitted (although it may have a high subjective probability that this is the case). Our results both show that there are previously unnoticed opportunities to optimize the protocol, and help to clarify what should be the specification of the protocol (the previous literature generally describes the protocol without providing a formal specification beyond the statement that it is intended for anonymous broadcast.) 
  
  The structure of the paper is as follows. We give a brief introduction to the logic of knowledge and epistemic model checking in Section~\ref{sec:kintro}. In Section~\ref{sec:kbp} we discuss knowledge-based programs and describe our methodology for the development of their implementations using epistemic model checking. The Dining Cryptographers problem and its extensions are introduced in Section~\ref{sec:dc}. In Section~\ref{sec:dckbp}, we describe the application of our methodology to this protocol. Finally, some conclusions are drawn in Section~\ref{sec:concl}.

\section{Model Checking Epistemic Logic} \label{sec:kintro} 

Epistemic logics are a class of modal logics that
include operators whose meaning concerns
the information available to agents in a 
distributed or multi-agent system. 
We describe here briefly a
version of such a logic combining operators for 
knowledge and linear time, and its semantics in 
a class of structures  known in the literature as  {\em interpreted 
systems} \cite{FHMVbook}. 
We then discuss the model checker MCK \cite{mck}, which is based on this semantics. 

Suppose that we are interested in systems comprised of $n$ agents
and a set $\Prop$ of atomic propositions. 
The syntax of the fragment of the logic of knowledge and time relevant for this paper 
is given by the following grammar: 
\[
\phi ::= \top \mid p \mid \neg \phi \mid \phi \wedge \phi \mid K_i\phi \mid \Next\phi 
\]
where $p\in \Prop$ is an atomic proposition and $i\in \{1\ldots n\}$ is an agent. 
(We freely use standard boolean operators that can be defined 
using the two given.) Intuitively, the meaning of $K_i\phi$ 
is that agent $i$ knows that $\phi$ is true, and $\Next\phi$ means that 
$\phi$ will be true at the next moment of time. 

The semantics we use is the  {\em interpreted systems} model 
for the logic of knowledge \cite{FHMVbook}. For each $i=0\ldots n$, 
let $S_i$ be a set of states. For $i=0$, we interpret $S_i$ 
as the set of possible states of the environment within which the agents operate; 
for $i=1\ldots n$ we interpret $S_i$ as the set of {\em local states} of agent $i$. 
Intuitively, a local state captures all the concrete pieces of information on the 
basis of which an agent determines what it knows. 
We define the set of {\em global state} based on such  collection of environment and 
local states, to be the set $S=S_0\times S_1\times \ldots \times S_n$. 
We write $s_i$ for the $i$-th component (counting from 0) of a global state $s$. 
A {\em run} over $S$ is a function $r:\nat \rightarrow S$. 
An {\em interpreted system} for $n$ agents is a tuple $\I = (\R,  \pi)$, 
where $\R$ is a set of runs over $S$, and $\pi: S\rightarrow \powerset{\Prop}$ is an interpretation function. 

A {\em point} of $\I$ is a pair $(r,m)$ where $r\in \R$ and $m\in \nat$. 
We say that two points  $(r,m),(r',m')$ are {\em indistinguishable} to agent $i$, and write $(r,m)\sim_i(r',m')$, 
if $r(m)_i = r'(m')_i$, 
i.e., if agent $i$ has the same local state at these two points. 
We define the semantics of the logic by means of a relation $\I,(r,m)\models \phi$, 
where $\I$ is an intepreted system, $(r,m)$ is a point of $\I$ and $\phi$ is a formula. 
This relation is defined inductively as
follows: 
\begin{itemize} 
\item $\I,(r,m) \models p$ if   $p\in \pi(r(m))$,  
\item 
$\I,(r,m)\models \neg \phi$ if not $\I,(r,m)\models \phi$
\item 
$\I,(r,m)\models \phi_1\lor \phi_2$ if $\I,(r,m)\models \phi_1$ or $\I,(r,m)\models \phi_2$ 
\item 
$\I,(r,m)\models \Next \phi$ if $\I,(r,m+1)\models \phi$

\item 
$\I,(r,m)\models K_{i} \phi$ if  for all  points $(r',m')$ of $\I$  such that $r(m)  \sim_i r'(m')$  we have $\I,(r',m') \models \phi$
\end{itemize} 
We note that the semantics of the knowledge operator depends not just on the 
run at which the formula is being evaluated, but also the set of all possible runs. 
Changing the set of runs (e.g., by 
making 
changes to the protocol), can change what an agent knows. 
Since knowledge-based programs change agent behaviours based on what the agent knows, 
this makes the semantics of knowledge-based programs somewhat subtle. 

MCK is a model checker based on this semantics for the logic of knowledge. 
For a given interpreted system $\I$, and a specification $\phi$ in the 
logic of knowledge and time, MCK computes whether 
$\I,(r,0) \models \phi$ holds for all runs $r$ of $\I$.  

Since interpreted systems are infinite structures, MCK 
allows an interpreted system to be given a finite description in the 
form of a program from which the interpreted 
system can be generated. This description is given using:  

\be
\item A list of global variables making up states of the 
environment, and their types. 

\item A listing of the agents in the system, together with the global 
variables that they are able to access. For each agent, we 
may also introduce local variables. If $v$ is a local variable of 
agent $A$, then we may refer to this variable in specification
formulas as $A.v$. Local variables may be aliased to global 
variables. 

A subset of the local  variables is specified as being {\em observable} to the agent. 
This means that it will be taken into account in the 
definition of the indistinguishability relation for the agent. 

\item A statement {\tt init\_cond $\phi$}, 
where $\phi$ is a boolean formula. All assignments satisfying 
this formula represent an initial state of the system.

\item A program that describes the protocol executed by each agent. 
The protocol describes how the agent chooses its actions 
depending on its history. 

\ee
Executing the agent protocols starting at an initial state
generates a set of runs, that we take to be the set of 
runs of the interpreted system generated by input script. 
(The agents operate in lock-step, each agent executing 
a single action in each step. Write-conflicts are syntactically prevented.) 
If $V$ is the set of all local and global variables in the system, 
then the component $s_0= r(n)_0$ of the global state at each point $(r,n)$
of a run $r$ is a well-typed  assignment of values to the variables $V$. 
The local state $s_i$ of agent $i$ in these runs are defined using the 
variables declared to be local. MCK allows this 
to be done in a number of ways, each giving a
different semantics for the knowledge operators. 
The construction of local states relevant to the present paper is the 
{\em perfect recall interpretation}. Writing $s_0\upharpoonright V_i$ for the 
restriction of the assignment $s_0$ to the variables $V_i$ 
that are observable to agent 
$i=1\ldots n$, 
the local states are defined to be the sequence
$$r(n)_i = (r(0)_0\upharpoonright V_i)\, (r(1)_0\upharpoonright V_i)\, \ldots (r(n)_0\upharpoonright V_i),$$
i.e., the local state is the history of all values of the variables observable to the agent. 

This perfect recall intepretation of knowledge is particularly relevant for analyses in which 
security or the optimal use of information are of concern. In 
both cases, we are interested in determining the maximal information 
that an agent is able to extract from  what it observes. Both issues are significant in 
the example that we study in this paper. MCK is the only model checker currently available  that 
supports symbolic model checking for the perfect recall interpretation of knowledge.

\section{Implementation of Knowledge-based Programs} \label{sec:kbp} 

Knowledge-based programs \cite{FHMVbook} are like standard programs, except that 
expressions may refer to agent's knowledge. 
That is, in a knowledge-based 
program for agent $i$, we may find 
statements 
 of the forms
$$\mbox{if $\phi$ then $P_1$ else $P_2$}$$
and $v:= \phi$, 
where $\phi$ is a formula of the logic of knowledge that is 
a boolean combination of atomic formulas concerning the agent's local variables
and formulas of the form $K_i\psi$, and $P_1,P_2$ are knowledge-based programs for agent $i$. 

Unlike standard programs, knowledge-based programs cannot in general be directly executed,
since, as noted above, the satisfaction of the knowledge subformulas depends on the
set of all runs of the program, which depends on the actions taken, which 
in turn depends on the satisfaction of these knowledge subformulas. 

This apparent circularity is handled by treating knowledge-based programs as specifications, and 
defining when a concrete standard program satisfies this specification. 
Suppose that we have a standard program $P$ of the same syntactic structure as the knowledge-based program ${\bf P}$, 
in which each knowledge-based expression 
$\phi$ is replaced by a concrete predicate $p_\phi$ of the local variables of the agent. 
In order to handle the perfect recall semantics, we  
also allow $P$ to add local {\em  history variables} and code fragments of the form $v:=e$, where $e$ is an expression, 
that update these history variables, so as to make information about past states available at the
current time.  The predicate $p_\phi$ may depend on the history variables. 

The concrete program $P$ generates a set of runs that we can take to be the basis of an interpreted system 
$\I(P)$. 
We now say that $P$ is an {\em implementation} of the knowledge-based program ${\bf P}$ if
for each formula $\phi$ in a conditional, we have that in the interpreted system 
$\I(P)$.  
the formula $p_\phi \dimp \phi$ is valid (at times when the condition is 
used).
That is, the
concrete condition is equivalent to the knowledge condition in the implementation. 
In general, knowledge-based programs may have no implementations, a behaviourally unique implementation, or
many implementations. Some conditions are known under which a behaviourally  unique implementation is 
guaranteed to exist. One of these conditions is  that agents have perfect recall and 
all knowledge formulas in the program refer to the present time (rather than to the past or future). 
This case will apply to the knowledge-based programs we consider in this paper, so we are 
guaranteed behaviourally unique implementations.  

We now describe a partially automated process, using epistemic model checking, 
that can be followed to find implementations of knowledge-based programs ${\bf P}$
(provide these terminate in a finitely bounded time: this applies to our examples)
The user begins by introducing a local boolean variable $v_\phi$ for each knowledge formula
$\phi= K_i\psi$ in the knowledge-based program, and replacing $\phi$ by $v_\phi$. 
Treating $v_\phi$  as a history variable, the user may also add to the program statements of the form $v_\phi:=e$, 
relying on their intuitions concerning situations under which the epistemic formula $\phi$ will be true. This produces a
standard program $P$ that is a candidate to be an implementation of the knowledge-based program
${\bf P}$. 
(It has, at least,  the correct syntactic structure.) 

To verify  the correctness of $P$ as an implementation of ${\bf P}$, the user must now 
check that the variables $v_\phi$ are being maintained so as to be equivalent to the 
knowledge formulas that they are intended to express. This can be done using epistemic model checking, 
where we verify formulas of the form 
$$X^n (pc_i=l \rimp (v_\phi \dimp K_i\psi) )$$
where $n$ is a time at which the test containing $\phi$ may be executed, 
$pc_i$ is the program counter of agent $i$ and $l$ is a label for the 
location of the 
expression
containing $\phi$. 
(This conditioning on the 
program counter 
can be dispensed with when the 
expression is known to always occur at particular times $n$, 
as it always is in our examples. 
More generally, 
we would write a formula that checks equivalence at {\em all} times for nonterminating 
programs, but the  resulting model checking problem is undecidable with respect to the perfect recall semantics.) 

In general, the user's guess concerning the concrete condition that is equivalent to the  knowledge formula 
may be incorrect, and the 
model checker will report the error. In this case, the model checker can be used  to generate an {\em error trace}, a 
partial run leading to a situation that falsifies the formula being checked. 
The next step of our process requires the user to analyse this error trace (by inspection and human reasoning) 
in order to understand the source of the error in their guess for the concrete condition representing the knowledge formula. 
As a result of this analysis, a correction of the assignment(s) to the variable $v_\phi$ is made 
by the user
(this step may require some ingenuity on the part of the user.) 
The model checker is then invoked again to check the new guess. 
This process is iterated until a guess is produced for which all the formulas of interest are found to be true, 
at which point an implementation of the knowledge-based program has been found. 

In many cases, this process can proceed monotonically. Starting from an initial 
assignment $v_\phi :=e$, where $e$ is a condition that the user can easily see to be {\em sufficient}  
for $K_i\psi$, the error trace leads to the identification of a situation where $i$ may know $\psi$, which is not covered by the 
condition $e$. (That is, where $K_i\psi \rimp e$ 
does not hold.) 
An analysis of this condition may lead to the 
discovery of another sufficient condition $e'$. In this case, the user can take as the next guess the 
assignment 
$v_\phi:=e\lor e'$. 
Continuing in this way, we obtaining an increasing sequence of concrete lower
approximations to the knowledge formula, eventually converging 
to
the correct implementation. 
(We note that such a condition $e'$ can always be found,  since we may always take it
to be a complete description of the run producing the counter-example. Finding a
good generalization that remains a sufficient condition for the
knowledge formula may be more difficult.) 

In general, monotonicity is not guaranteed, but it  obtains in our example in this paper. 
We leave the question of characterizing the situations where monotonicity
applies to future work, and turn to a demonstration of the process on a particular
example, introduced in the next section. 

\section{Chaum's dining cryptographers protocol}\label{sec:dc} 

Chaum's dining cryptographers protocol \cite[p. 65]{chaum} is an example of a 
protocol for secure multiparty computation: it enables 
the value of a function of a group of agents to be computed
while revealing nothing more than that value. 
Chaum introduces the protocol  with the following story: 

\begin{quote}
Three cryptographers are sitting down to dinner at their favourite restaurant.
Their waiter informs them that arrangements have been made with the maitre d'hotel
for the bill to be paid anonymously. One of the cryptographers might be paying for
the dinner, or it might have been NSA (U.S National Security Agency). The three
cryptographers respect each other's right to make an anonymous payment, but they
wonder if NSA is paying. They resolve their uncertainty fairly by carrying out the following protocol:\\

Each cryptographer flips an unbiased coin behind his menu, between
him and the cryptographer on his right, so that only the two of
them can see the outcome. Each cryptographer then states aloud
whether the two coins he can see -- the one he flipped and the one
his left-hand neighbor flipped--fell on the same side or on
different sides. If one of the differences uttered at the table
indicates that a cryptographer is paying; an even number indicates
that NSA is paying (assuming that the dinner was paid for only
once). Yet if a cryptographer is paying, neither of the other two
learns anything from the utterances about which cryptographer it
is.
\end{quote}

This version of the dining cryptographers protocol has frequently been the focus
of studies of verification of security protocols, but it is just one of many variants 
discussed in Chaum's paper. One of Chaum's considerations is the use of the protocol for
more general anonymous broadcast applications, and he writes: 
\begin{quote} 
The cryptographers become intrigued with the ability to make messages
public untraceably.  They devise a way to do this at the table for a
statement of arbitrary lenght: the basic protocol is repeated over and
over; when one cryptographer wishes to make a message public, he merely
begins inverting his statements in those rounds corresponding to 1's in
a binary coded version of his message. If he notices that his message
would collide with some other message, he may for example wait for a
number of rounds chosen at random from some suitable distribution
before trying to transmit again.
\end{quote}  
He notes that ``undetected collision results only from an odd number
of synchronized identical message segments''.  As a particular
realization of this idea, he discusses grouping communication into blocks and 
the use of the following 
{\em 2-phase broadcast} protocol using {\em slot-reservation}:  
\begin{quote} 
In a network with many messages per block, a first block may be used by various anonymous senders 
to request a ``slot reservation'' in a second block. A simple scheme would be for each 
anonymous sender to invert one randomly selected bit in the first block for each slot they wish to reserve
in the second block. After the result of the first block becomes known, the participant 
who caused the ith bit in the first block sends in the ith slot of the second block. 
\end{quote} 
This idea has been implemented as part of the Herbivore
system\cite{Herbivore}.  (Herbivore also adds mechanisms for dividing
the group of participants into cliques of sufficient size to provide
reasonable anonymity guarantees, as well as protocols for joining a
leaving the group of particpants - we will not discuss these extension
here.) The Herbivore authors note that  
\begin{quote} 
If an even number of nodes attempt to reserve a given slot, the
collision will be evident in the reservation phase, and they will
simply wait until the next round to transmit. If an odd number of
nodes collide, the collission will occur during the 
transmission phase.
\end{quote} 

The remarks above do not constitute a concrete definition of the
protocol, and leave a number of questions concerning the
implementation open.  For example, what exact test is applied to
determine whether there is a collision? 
Which agents are able to detect a collision? 
Are there situations where some agent expects
to receive a message, but a collision occurs that it 
does not detect (although some other agent may do so?) 

Note that each round of the DC protocol has been proved correct, but
what about the way in which the rounds are combined? It is not
immediately clear that there are not subtle flows of information!

Prior knowledge of the participants may also affect the 
flow of information. For example, suppose that the protocol is being used for the
participants in a referendum to anonymously announce their votes. In
this case it is known that all particpants will attempt to reseve a
slot - does this information change the flow of information in any
way? If so, does it affect the security of the protocol?
One of the benefits of verification by 
epistemic 
model checking is that 
it permits such questions about variants of a protocol, 
and its application in a particular setting to be investigated 
efficiently without requiring reconstruction of possibly complex proofs.

\section{The 2-phase Broadcast Protocol as a Knowledge-based Program}\label{sec:dckbp}

It is interesting to note that the descriptions of the 2-phase protocol
above are, in their level of abstraction, 
more like knowledge-based programs than like concrete
implementations. In this section, we explicitly study the 
protocol from this perspective, and apply our partially automated 
methodology to derive the concrete implementations. 
We consider a setting with 3 agents who use 3 slots for their broadcast. 
Each slot permits the transmission of a single-bit message. 

\subsection{The Knowledge-Based Program} 

Figure~\ref{fig:kbp} represents the 2-phase protocol as 
 a knowledge-based program. The parameters of the protocol in the 
 first line alias certain local variables to global variables in the environment.  
Variable \verb+i+ is a number  
in the range 1..3 
used to index the present instance of the protocol, and variables 
\verb+keyleft+ and \verb+keyright+ represent keybits (referred to as ``coins", above), 
which are shared between by agents in the appropriate pattern. 
Note that since a fresh set of keybits needs to be used for each 
instance of the basic Dining Cryptographers protocol (which we run 6 times here), we assume that an external process 
generates fresh values for these keybit variables at each step; we omit the details. 
The final variable \verb+said+ in the parameters represent the
array of public announcements by the agents at each step.  
All arrays are assumed to be indexed starting from 1. 
The local variable \verb+slot-request+ records the slot number 
(in the range 1..3) 
that this 
agent will attempt to reserve. If \verb+slot-request+=0, then the 
agent will not attempt to reserve any slot.  The variable
\verb+message+ records the single bit message that the agent 
wishes to anonymously broadcast (if any). 
Variables for which an initial value is not explicitly specified can take any initial value. 
We write `$\oplus$' for the exclusive or operation. 

\begin{figure} [h]
{\bf protocol} dc\_agent(i:[1,3], \keyleft,\keyright,\said[3]:Bool) \{\\
local variables:\\ 
\hspace*{20pt}  \slotrequest:[0,3],\\
\hspace*{20pt}  \msg:Bool,\\
\hspace*{20pt}  \rcvd0[3], \rcvd1[3], \dlvrd : Bool (initially false);\\
//reservation phase\\
\verb+for+ ($s =1$; $s \leq 3$; $s$++)\\ 
  $\{$\\
\hspace*{5pt}   \said[i] := (\keyleft $\oplus$ \keyright $\oplus$ (\slotrequest=$s$));\\
  $\}$ \\
//transmission phase\\
\verb+for+ ($s =1$; $s \leq 3$; $s$++) \\
  $\{$ \\
\hspace*{5pt} \verb+if+ ($\slotrequest=s$ $\land$ $\neg K _{i}(\conflict(s))$\\
\hspace*{20pt} \verb+then+ \said[i] := (\keyleft $\oplus$ \keyright $\oplus$ \msg) \\
\hspace*{20pt} \verb+else+ \said[i] := (\keyleft $\oplus$ \keyright $\oplus$ false);\\
\hspace*{5pt} \rcvd0[s] := $K_i(\sender(i,0,s))$; \\ 
\hspace*{5pt} \rcvd1[s] := $K_i(\sender(i,1,s))$ \\ 
  $\}$; \\
\dlvrd := $
\begin{array}[t]{r}
\bigwedge_{x\in Bool, t=1..3 } ((\mbox{$\msg = x$} ~ \land ~\mbox{$\slotrequest = t$}) \rimp\\ 
 K_i ( \bigwedge_{j\neq i} K_j\sender(j,x,t)))
 \end{array}
 $\\
\}
\textbf{ \caption{ \label{fig:kbp}The knowledge-based program $CDC$}}
\end{figure}

The term $\conflict(s)$ in the knowledge-based program 
represents that there is a conflict on slot $s$. 
This is a global condition that is defined as
$$ \conflict(s) = \bigvee_{i\neq j} (i\verb+.slot-request+= s= j\verb+.slot-request+)~.$$
i.e., there exist two distinct agents $i$ and $j$ both requesting slot $s$. 

The term $\sender(i,x,s)$ represents that an agent other than $i$ 
is sending message $x$ in slot $s$; this is defined as 
$$ \sender(i,x,s) = \bigvee_{j\neq i} (j.{\tt message} = x \land j.\slotrequest =s)~.$$
Thus the variable \verb+rcvd0[s]+ is assigned to be true if in round $s$, 
the agent  learns that someone else is trying to send the bit 0, and
similarly for \verb+rcvd1[s]+. 
This addresses an issue that is not explicitly mentioned in the discussion of the
two-phase protocol above, viz., how does an agent know whether it has
received a transmission from another? Note that this is pertinent because the 
knowledge-based program allows that, although an agent has declared that it 
wishes to reserve a slot, it may still back off from the transmission if
it discovers that there is a conflict. But will the receiver always know that it has done so? 

We note that this representation of the 2-phase protocol as a 
knowledge-based program 
is {\em speculative}: an agent transmits in a slot so long as it does not know 
that there is a conflict. This allows that a collision will occur during the transmission phase. 
One of the benefits of the knowledge-based approach is that it makes 
explicit the difference between this and another interpretation of the protocol, where in 
place of the condition $\neg K _{i}(\conflict(s)$) we use the condition $K _{i}(\neg \conflict(s)$). 
In this {\em conservative} version, an agent would broadcast only  if it is certain that there is not
a conflict on its desired slot. Both versions may be appropriate depending on the 
circumstances, but we focus our discussion here on the speculative version.    

Since an agent may attempt to reserve a slot, and then back off, 
or may send in a reserved slot without success, the protocol does not 
guarantee that the message will be delivered. In this case, the 
agent is required to retry the transmission in the next run of the protocol. 
So 
that 
it can determine whether a retry is necessary, the final assignment 
to the variable $\dlvrd$ captures  whether the agent knows that 
its (anonymous) transmission has been successful. This is the case if
all other agents know that  {\em some}  agent sent the bit $i.\msg$ in 
slot $j.\slotrequest$. (Subtleties about the semantics of the logic of
knowledge prevent simplification of this formula by substitution of 
these expressions for $x$ and $t$.) 

In order to set up the appropriate configuration of the 3 agents and to alias their parameters
to variables in the environment, we use the following declaration block:
\vspace{-20pt}
 \begin{figure}[H]
\begin{verbatim} 
agent C2 : dc_agent(1,k31,k12,said) 
agent C3 : dc_agent(2,k12,k23,said) 
agent C3 : dc_agent(3,k23,k31,said) 
\end{verbatim}
\vspace{-10pt} 
\end{figure}
\vspace{-25pt} 
\noindent
where the \verb+k+$ij$ are boolean variables that 
represent the keybit shared between agent $i$ and agent $j$. 

In Figure~\ref{fig:imp}, we give the generic structure of a possible implementation of the 
knowledge-based program, as we seek using our partially-automated process. 
The lines marked with (+) indicate places of difference with CDC. 

Here we have introduced some history variables 
\verb+rr[s]+ 
that record 
the {\em round results} \verb+said[0]+$\oplus$ \verb+said[1]+$\oplus$\verb+said[2]+ 
obtained from each round 
$s$ 
of the basic Dining Cryptographers protocol. 
Note that, because of the pattern of sharing of the 
keybits between  the agents, this expression contains each  
keybit value twice, so that the keybits cancel out, leaving just the 
exclusive-or of the actual content being transmitted by each of the agents
(in each assignment to \verb+said[i]+, this is the final term in the exclusive-or). 
In particular, under the assumption that just one agent has a genuine message $x$ to transmit in round $j$, 
and the others transmit $false$, we obtain that
 \verb+rr[j]+=$x$.

The variable \verb+kc[s]+ is used to represent the epistemic condition
concerning conflict 
 in the 
knowledge-based program ($\neg K _{i}(\conflict(s))$ or $K _{i}(\neg \conflict(s))$, 
depending on whether we are dealing with the speculative or the conservative version). 
Thus, in verifying that we have an implementation, the 
key condition to be checked is whether $\verb+kc[s]+\dimp \neg K _{i}(\conflict(s))$ 
(respectively, $\verb+kc[s]+ \dimp K _{i}( \neg \conflict(s))$) is valid at the times the \verb+if+ statement is executed. 
The main difficulty in finding an implementation is to find the appropriate 
concrete 
assignment
for this variable that will make this condition valid. 
Similarly we seek assignments to the variables \verb+rcvd0[s], recvd1[s]+ that give these the intended meaning.

\begin{figure} [h]
{\bf protocol} dc\_agent(i:[0,2], \keyleft,\keyright,\said[3]:Bool) \{ \\
local variables:\\ 
\hspace*{20pt}  \slotrequest:[0,3],\\
\hspace*{20pt}  \msg:Bool,\\
\hspace*{20pt}  \rcvd0[3], \rcvd1[3]:Bool (initially false),\\
\hspace*{20pt}  \rr[6]:Bool,\hfill(+)\\
\hspace*{20pt}  \kc[3]:Bool (initially false);\hfill(+)\\
//reservation phase\\
\verb+for+ ($s =1$; $s \leq 3$; $s$++)\\ 
  $\{$\\
  \hspace*{5pt}   \said[i] := (\keyleft $\oplus$ \keyright $\oplus$ (\slotrequest == $s$)); \\
\hspace*{5pt}    
\rr[s] 
:=\said[0]$\oplus$ \said[1]$\oplus$ \said[2]; \hfill(+)\\
  $\}$ \\
//transmission phase\\
\verb+for+ ($s =1$; $s \leq 3$; $s$++) \\
  $\{$ \\
\hspace*{5pt} \kc[s] :=???;\hfill(+)\\
\hspace*{5pt} \verb+if+ (\slotrequest == s $\land$ \kc[s])\\
\hspace*{20pt}   \verb+then+ \said[i] := (\keyleft $\oplus$ \keyright $\oplus$ \msg) \\
\hspace*{20pt}   \verb+else+ \said[i] := (\keyleft $\oplus$ \keyright $\oplus$ false);\\
\hspace*{5pt}   rr[s+3] := \said[0]$\oplus$ \said[1]$\oplus$ \said[2]; \hfill(+)\\
   \hspace*{5pt} \rcvd0[s] := ???;\hfill (+) \\ 
   \hspace*{5pt} \rcvd1[s] := ???;\hfill (+)\\ 
  $\}$\\
  \dlvrd := ??? \hfill (+)\\ 
  \}
\textbf{ \caption{\label{fig:imp}A generic implementation of $CDC$}}
\end{figure}
\vspace{-20pt}

\subsection{Verification Conditions} 

In order to apply our methodology, it is necessary for the user to 
substitute a guess for parts of the implementation marked `???', 
and then to use model checking to check the correctness of the 
guess. We now discuss the formulas that are used to verify the implementation.  
In general, the conditions need to be verified only at specific times $n$, straightforwardly 
determined from the structure of the program. We generally omit discussion of this. 

The first formula of interest concerns the correctness of the 
guess for the knowledge condition $\neg K_i(\conflict(s))$ (in case of the 
speculative implementation, or $K_i(\neg \conflict(s))$ (in the case of the 
conservative implementation). In the implementation, 
this condition is represented by the variable \verb+kc[s]+. 

\textit{Specification 1:
{\tt kc[s]}  correctly represents knowledge of the existence of a conflict in slot $s=1..3$. }
In case of the speculative interpretation, we use the formula

$$X^n  (i.\verb+kc[s]+ \dimp \neg K_{i} (\conflict(s)))~~~~~~~(1s)$$
and in case of the conservative implementation, we use the 
formula 
$$X^n  (i.\verb+kc[s]+ \dimp  K_{i} (\neg \conflict(s)))~~~~~~~~(1c)$$
(In both cases, the appropriate values of $n$ are 7, 12 and 17, 
where we treat the {\tt for}  loops as macros and the {\tt if} conditions 
as taking zero time.)  

As remarked above, it has been claimed that the 2-phase protocol 
is guaranteed to detect a conflict either in the  slot-reservation phase
or else in the transmission phase. To verify this, we can  
use the following specification: 

\textit{Specification 2: A conflict is always detected. } 
$$X^n  (\conflict(s) \rimp  K_{i} (\conflict(s)))$$
where we may take time $n$ to correspond to the final 
time in the protocol. We remark that the converse implication is trivial 
from the semantics of knowledge.

As will discuss below, Specification 2 is arguably too strong, 
since agents may not be able to learn about conflicts on 
slots they do not reserve. Thus, the following weaker 
specification is also of interest. 

\textit{Specification 3: If there is a slot conflict involving agent 
$i$, then agent $i$ detects it.}
$$ X^n((\conflict(s) \land i.\slotrequest = s) \rimp K_i(\conflict(s)))$$
where again we take $n$ to correspond to the end of the protocol.

Next, the protocol has some positive goals, viz., to allow agents to broadcast some 
information, and to do so anonymously. Successful reception of a bit by the time $n$ immediately after the 
transmission in slot $s$  is 
intended to be represented by the variables \verb+rcvd0[s]+ and \verb+rcvd1[s]+. 
To ensure that the assignments to these variables correctly implement their
intended meaning in the knowledge-based program, we use specifications of the
following form. 

\textit{Specification 4: reception variables  correctly represent transmissions by others} 
$$ X^n(i.\rcvd0[s] \dimp K_i(\sender(i,0,s))) ~~~~~~~~~(4a)$$ 
and 
$$ X^n(\rcvd1[s] \dimp K_i(\sender(i,1,s))) ~~~~~~~~~ (4b)$$ 

Similarly, we need to verify correct implementation of the agent's knowledge about whether its transmission is 
successful.  

\textit{Specification 5: delivery variables  correctly represent knowledge about delivery} 
$$
\begin{array}{l} 
X^n ( i.\dlvrd \dimp 
\bigwedge_{x\in Bool, t=1..3 } (i.\msg = x  \land i.\slotrequest = t\\
~~~~~~~~~~~~~~~~~~~~~~~~~ \rimp K_i ( \bigwedge_{j\neq i} K_j\sender(j,x,t))))
\end{array} 
$$

Finally, the aim of the protocol is to ensure that when information is transmitted, this 
is done anonymously. An agent may know that one of the other two agents 
has a particular message value, but it may not know what that value is
for a specific agent. We may write the fact that agent $i$ knows the 
value of a boolean variable $x$ by the notation $\hat{K}_i(x)$, defined 
by 
$$\hat{K}_i(x) = K_i(x) \lor K_i(\neg x)~.$$
Using this, we might first attempt to specify anonymity as
$\bigwedge_{j\neq i} ( \neg \hat{K}_i(j.\msg)$, i.e., agent $i$ knows no other's message. 
Unfortunately, the protocol cannot be expected to satisfy this: suppose that 
all agents manage to broadcast their message and all messages have the same value
$x$: then each knows that the other's value is $x$. We therefore write the 
following weaker specification of anonymity: 

\textit{Specification 6:  The  protocol preserves anonymity}
$$ X^n ( \bigvee_{x=0,1} K_i( \bigwedge_{j\neq i} (j.\msg = x)) \lor 
\bigwedge_{j\neq i} ( \neg \hat{K}_i(j.\msg)))$$ 
to be evaluated with $n$ set to the final time of the protocol. 

\subsection{Finding an implemention of the knowledge-based program} 

We now illustrate how we find an implementation of the knowledge-based program 
using our methodology. We focus here on the speculative version, and consider a 
scenario where 
the number of agents that are seeking to broadcast$-$ is initially unknown, 
and could be any value from the set \{$0..3$\}. 

 Our first task in implementing the knowledge-based program is to find an 
 appropriate assignment for the variables $\kc[s]$, and to verify that this 
 assignment correctly represents knowledge about
slot conflicts and  validates {\it Specification 1}. It is plain from the 
 discussion above that if an agent attempts to reserve slot 
 $s$, but sees that the round result for that reservation attempt is
 not $true$, then this must be because some other agent 
 also attempted to reserve the slot. Thus, in this case the agent 
 detects a conflict. A reasonable guess for the 
 assignment to $\kc[s]$ to represent $\neg K_i(\conflict(s))$ is therefore 
 $$ \kc[s] := \neg (\slotrequest =s \land  \neg \rr[s] = false)~. $$ 
Indeed, this proves to be the correct choice: if we now model check {\it Specification 1s} then 
we find that this specification is true.%
\footnote{Strictly, in order to model check this claim, we 
first need to fill in the other `???' assignments. 
We remark that because of independencies, the outcome of model checking 
{\it Specification 1s} is the same  {\em whatever} we choose for the other
`???' assignments. We omit a detailed argument for this here.}

The next question of interest is then whether {\it Specification 2} holds , 
as claimed. The answer obtained by model checking 
is that it does not, and the counter-example discovered is the following: 

\textbf{Example 1: (None of the agents discover conflict)}  
Suppose that all agents (C1, C2, C3) would like to reserve slot 
2 and  each has message $\true$. 
The round results $\rr[s]$ are shown in 
on the left in Figure~\ref{fig:cex},  
where we show for each agent the contribution 
other than keybits (which cancel out).    
\begin{figure} [h]
\centering
\begin{tabular}{ccc}
\begin{tabular} [h] {|l|c|c|c||c|c|c|}
\hline
$s$ & 1 & 2 & 3 & 4 & 5 & 6\\
\hline
Agent C1 & 0 & 1 & 0 & 0 & 1 & 0 \\
\hline
Agent C2 & 0 & 1 & 0 & 0 & 1& 0 \\
\hline
Agent C3 & 0 & 1 & 0 & 0 & 1& 0 \\
\hline
\hline
$\rr[s]$ & 0 & 1 & 0 & 0 & 1 & 0 \\
\hline
\end{tabular}
& ~~~~~~~~~~~~~~ & 
\begin{tabular} [h] {|l|c|c|c||c|c|c|}
\hline
$s$ & 1 & 2 & 3 & 4 & 5 & 6\\
\hline
Agent C1 & 0 & 1 & 0 & 0 & 1 & 0 \\
\hline
Agent C2 & 0 & 0 & 0 & 0 & 0 & 0 \\
\hline
Agent C3 & 0 & 0 & 0 & 0 & 0 & 0 \\
\hline
\hline
$\rr[s]$ & 0 & 1 & 0 & 0 & 1 & 0 \\
\hline
\end{tabular}

\\
~\\[-5pt]

$\slotrequest = [2,2,2]$,
& & $\slotrequest = [2,0,0]$ \\ 
 $\msg = [\true,\true,\true]$ & & $\msg = [\true,\true,\true]$
 
 \end{tabular} 
\caption{Runs indistinguishable to C1\label{fig:cex} } 
\end{figure}

Now from agent $C1$'s perspective, this run of the protocol is indistinguishable from  
another run where only $C1$ attempts to reserve slot 2, and it still has message $\true$, shown 
on the right in Figure~\ref{fig:cex}. 
Hence we have a situation where although there is a conflict agent $C1$ 
cannot know that there is a conflict, and {\it Specification 2} fails.%
\footnote{In fairness to the authors of \cite{Herbivore}, they state  
that messages are sent with an MD5 checksum, so most conflicts of 
messages somewhat longer than a single bit 
 would in fact be detected with high probability through corruption of this checksum. However, even with this device, 
 collisions of 3 identical messages would still go undetected, 
 as noted by Chaum.  Our example shows that the appropriate formalization of this claim should be 
 probabilistic, something that we do not take up here.}  
 Indeed, we see that the more liberal {\it Specification 3} 
 also fails in this example.   
 
In the discussion above, we have focussed on the agent's knowledge that there is a conflict.  
From the point of view of determining the appropriate assignments to 
the variables $\rcvd0$ and $\rcvd1$, it would be helpful to 
determine under what circumstances an agent knows that there will be a transmission on a 
slot but there is {\em not} a conflict on that slot.  Thus, it would be helpful to have a predicate 
$i.\conflictfree(s)$
that is equivalent to $K_i(\bigvee_j j.\slotrequest = s \land \neg \conflict(s))$. We now investigate this question, 
and use it to illustrate the iterative procedure to obtain local predicates that are 
equivalent to knowledge formulas. 

Plainly, a round-result of $\true$ during the reservation phase implies that someone 
wishes to send in that slot. However,  Example 1 also shows that 
$K_i\neg \conflict(s)$ cannot hold in case agent $i$ obtains 
round result $\true$ in a slot it intends to transmit in, and $\false$ in all 
other slots, since it is possible that all agents are attempting to 
transmit in the same slot. Hence a reasonable guess is 
$$ \conflictfree1(s) = \rr[s] = \true \land \neg (\wedge_{t\in \{1,2,3\} \setminus \{s\}} \rr[t] = \false)~.$$ 
When we model check $$X^n( i.\conflictfree1(s) \dimp K_i(\bigvee_j j.\slotrequest = s \land \neg \conflict(s))$$ at time $n$ after the 
transmission phase, we find that this formula is false. A counter-example
produced by the model checker shows that this happens when 
$C1$ and $C3$ request slot 3, and $C2$ requests slot $1$. 
Note that in this case the reservation round results are $(1,0,0)$. 
Here $C1$ and $C3$ detect a conflict in slot 3. Since there are 
only three agents, they are able to reason that the conflict must 
have been 2-way (else we have the scenario of Example 1). 
This means that they are able to deduce that there is 
{\em not} a conflict in slot 1. 

This example motivates a second guess for the predicate
$\conflictfree(s)$, viz., (when all variables are local to agent $i$)
$$
\begin{array}[t]{l}
 \conflictfree2(s)  =   \conflictfree1(s)~ \lor \\
~~~~ ( \rr[s] = \true~ \land~ \slotrequest \in \{1,2,3\} \setminus \{s\} ~ \land~ \rr[i.\slotrequest] =\false)~.
\end{array} 
 $$
Model checking this predicate for equivalence to $K_i(\bigvee_j j.\slotrequest = s \land \neg \conflict(s)),$
we still find that the equivalence does not hold. The counter-example produced this time 
is the situation where agents $C1$ and $C2$ do not request a slot, but 
agent $C3$ requests slot $s$ so that the round result of slot $s$ is $\true$.
Note that here, agents $C1$ and $C2$ know
that any slot collision must be 2-way, since they cannot be a participant. 
Since the reservation request on slot $s$ gave round result $1$, there must be 
exactly one agent requesting slot $s$. With some reflection, we note that 
agent $C1$ would have been able to draw the same conclusion about slots $2$ and $3$ 
in case the round result pattern were $(0,1,1)$. 
Thus, we are led to the following improved guess: 
$$
\begin{array}[t]{l}
 \conflictfree3(s) =  \conflictfree2(s) \lor 
( \rr[s] = \true \land \slotrequest \neq s ) 
\end{array} 
 $$
At this point, model checking shows that we have found the 
predicate we seek.

Returning now to the question of when agents learn the bit that another agent is transmitting, we guess the
assignment 
$$\rcvd1[s] := \rr[s] = \true \land \conflictfree3(s) \land \slotrequest \neq s~. $$
That is, the agent sees that there will be a conflict free transmission on slot $s$, 
but it is not itself using that slot. We now  model check {Specification 4b}. Somewhat surprisingly, this specification 
turns out to be false! The counter example returned is one in which 
the agent is $C1$, all agents reserve slot $1$, and the agents have messages $(1,1,0)$. 
Note that here,  the round result obtained for the transmission is $0$, so agent $C1$ detects
the collision, which it knows must have been 3-way. It can also reason that the other agents cannot both 
have had messages $\false$, since this would have produced 
round result $\false$,
thus, at least one must have had message $\true$!
This observation leads to the revised guess
$$
\begin{array}{l} 
\rcvd1[s] := (\rr[s] = \true \land \conflictfree3(s) \land \slotrequest \neq s) \lor \\ 
~~( \slotrequest = \true \land \rr[s+3] \neq \msg \land \bigwedge_{t\in \{1,2,3\} \setminus \{s\}} \rr[t] = \false)  ~.
\end{array}
 $$
We now find that {Specification 4b} holds, so we have correctly implemented
this part of the knowledge-based program.  
A similar assignment works for the assignment to $\rcvd0$
 and {\it Specification 4a}. 

This process can also be carried out also for the final specification 
{\it Specification 5}, which concerns the circumstances under
which a sender knows that their message has been received
by the others. One obvious situation when this is the case 
is when the sender $i$ knows that the slot on which they are
sending is conflict-free. Recall that this occurs only when two or more of the reservation round  
results equal $\true$, and note that this implies that all other agents also
know that the slot on which $i$ is sending is conflict-free. Thus
the others will receive that message that $i$ is sending (anonymously)
on this slot. This suggests the assignment
$$ \dlvrd := \slotrequest = 0 \lor \bigvee_{s\in \{1,2,3\}}  
\slotrequest = s \land \conflictfree3(s)~.$$
When we model check this with respect to {\it Specification 5}, we
find that that the specification holds, and we have a complete
implementation of the knowledge-based program.
Finally, we may also model check {\it Specification 6} 
and verify that the protocol preserves anonymity in the appropriate sense. This proves to be the case.

\section{Conclusion} \label{sec:concl}
We have demonstrated the application of our partially automated methodology 
for knowledge-based program implementation on a protocol for anonymous broadcast. 
While, like related studies \cite{HO03,HS04,MS,PS99,RS00,SA96}, we verify that an anonymity property  holds, 
the focus of our effort lies in other aspects of the protocol.

One of the main outcomes of the analysis is that the flows of information in the 
protocol are considerably more subtle than one might have expected. 
In particular, we find that there are circumstances, 
that go beyond those that have been identified in the literature, 
where  agents are able to obtain knowledge of each other's bits. 
Significantly, we make this discovery not manually, but using automated support. 
We also address in our work a number of questions that have not been 
considered in the prior literature, viz., under what circumstances can a receiver
be confident that they are receiving a transmission, and under what circumstances 
a sender can know that its transmission has been successful, and find complete 
answers to these questions in a particular scenario. 

On the other hand, being based on model checking of a concrete 
model under very particular assumptions, our approach lacks generality: 
it does not yield an immediate answer to how our conclusions are 
affected by changing the number of agents, their topology, 
or the initial assumptions concerning the number of agents wishing to transmit. 
However, the methodology provides an efficient means to 
experiment with such questions. 
We are presently investigating further variants using our 
methodology,  in order to obtain an empirical basis 
from which theoretical results may be generalized. 
Our present models are also starting to press the 
limits of the model checking technology (run times of the order of
hours for some queries, for protocols of around 20 steps), so we are 
also investigating optimizations that will increase the scale and 
complexity of the problems we can address. We plan to 
report on this in future work. 

\bibliographystyle{plain}
\bibliography{references}

\begin{thebibliography}{10}

\bibitem{BaukusM04}
Kai Baukus and Ron van~der Meyden.
\newblock A knowledge based analysis of cache coherence.
\newblock In {\em 6th Int. Conf. on Formal Engineering Methods}, volume 3308 of
  {\em LNCS}, pages 99--114. Springer, 2004.

\bibitem{chaum}
Chaum.
\newblock The dining cryptographers problem: Unconditional sender and recipient
  untraceability.
\newblock {\em Journal of cryptology}, pages 65--75, 1988.

\bibitem{DM86}
C.~Dwork and Y.~Moses.
\newblock {Knowledge and common knowledge in a Byzantine environment : crash
  failures}.
\newblock In {\em Proceedings of the 1986 Conference on Theoretical aspects of
  reasoning about knowledge}, pages 149--169, San Francisco, CA, USA, 1986.
  Morgan Kaufmann Publishers Inc.

\bibitem{DM90}
C.~Dwork and Y.~Moses.
\newblock Knowledge and common knowledge in a {B}yzantine environment: crash
  failures.
\newblock {\em Information and Computation}, 88(2):156--186, 1990.

\bibitem{FHMVbook}
R.~Fagin, J.~Y. Halpern, Y.~Moses, and M.~Y. Vardi.
\newblock {\em Reasoning about Knowledge}.
\newblock MIT Press, Cambridge, Mass., 1995.

\bibitem{mck}
P.~Gammie and R.~van~der Meyden.
\newblock {MCK: Model checking the logic of knowledge}.
\newblock In {\em Proceeding of the 16th Int. Conf. on computer Science Aided
  Verification (CAV'04)}, volume 3114 of LNCS, pages 479--483. Springer-Verlag,
  2004.

\bibitem{Herbivore}
S.~Goel, M.~Robson, M.~Polte, and E.~Sirer.
\newblock {Herbivore: A Scalable and Efficient Protocol for Anonymous
  Communication}.
\newblock Technical report, Cornell University, Ithaca, NY, February 2003.

\bibitem{Had87}
V.~Hadzilacos.
\newblock A knowledge-theoretic analysis of atomic commitment protocols.
\newblock In {\em PODS '87: Proceedings of the sixth ACM SIGACT-SIGMOD-SIGART
  symposium on Principles of database systems}, pages 129--134, New York, NY,
  USA, 1987. ACM.

\bibitem{HZ92}
J.~Y. Halpern and L.~D. Zuck.
\newblock A little knowledge goes a long way: knowledge-based derivations and
  correctness proofs for a family of protocols.
\newblock {\em Journal of the ACM}, 39(3):449--478, 1992.

\bibitem{HO03}
Joseph~Y. Halpern and Kevin~R. O'Neill.
\newblock Anonymity and information hiding in multiagent systems.
\newblock In {\em Proc. of the 16th IEEE Computer Security Foundations
  Workshop}, pages 75--88, 2003.

\bibitem{HS04}
Dominic Hughes and Vitaly Shmatikov.
\newblock Information hiding, anonymity and privacy: a modular approach.
\newblock {\em Journal of Computer Security}, 12 (1):3--36, 2004.

\bibitem{CGP99}
E.~M.~Clarke Jr., O.~Grumberg, and D.~A. Peled.
\newblock {\em Model Checking}.
\newblock The MIT Press, 1999.

\bibitem{mcmas}
A.~Lomuscio, H.~Qu, and F.~Raimondi.
\newblock {MCMAS}: A model checker for the verification of multi-agent systems.
\newblock In {\em CAV}, volume 5643 of {\em Lecture Notes in Computer Science},
  pages 682--688. Springer, 2009.

\bibitem{RS00}
P.~Ryan and S.~Schneider.
\newblock {\em The modelling and analysis of security protocols: the {CSP}
  approach}.
\newblock Addison-Wesley Professional, 2000.

\bibitem{SA96}
Steve Schneider and Abraham Sidiropoulos.
\newblock {{CSP} and anonymity}.
\newblock In {\em {Proc. of the European Symposium on Research in Computer
  Security (ESORICS)}}, pages 198--218. Springer-Verlag, 1996.

\bibitem{PS99}
Paul Syverson and Stuart Stubblebine.
\newblock Group principals and the formalization of anonymity.
\newblock In {\em FM '99: Proceedings of the Wold Congress on Formal Methods in
  the Development of Computing Systems-Volume I}, pages 814--833, London, UK,
  1999. Springer-Verlag.

\bibitem{MS}
Ron van~der Meyden and Kaile Su.
\newblock Symbolic model checking the knowledge of the dining cryptographers.
\newblock In {\em Proceedings of the 17th IEEE Computer Security Foundation
  Workshop}, pages 280--291. IEEE Computer Society, 2004.

\bibitem{demo}
J.\ van Eijck.
\newblock Dynamic epistemic modelling.
\newblock Technical report, Centrum voor Wiskunde en Informatica, Amsterdam,
  2004.
\newblock CWI Report SEN-E0424.

\end{thebibliography}

\end{document}